\documentclass[aps,prl,twocolumn,showpacs,10pt,superscriptaddress,preprintnumbers,nofootinbib]{revtex4-1}
\usepackage{epsfig,amssymb,amsmath,psfrag,epstopdf,color}
\pdfoutput=1
\usepackage{graphicx}
\usepackage{subfig}
\usepackage[margin=7pt,justification=centerlast]{caption}

\usepackage{paralist}
\usepackage{hyperref}

\allowdisplaybreaks

\def\beq{\begin{equation}}
\def\eeq{\end{equation}}
\def\bsp#1\esp{\begin{split}#1\end{split}}
\newcommand{\be}{\begin{equation}}
\newcommand{\ee}{\end{equation}}
\newcommand{\bea}{\begin{eqnarray}}
\newcommand{\eea}{\end{eqnarray}}

\def\Fig#1{Fig.~{\ref{#1}}}

\def\cN{{\mathcal N}}
\def\cO{{\mathcal O}}

\def\to{\rightarrow}

\def\ksl{\not{\hbox{\kern-2.3pt $k$}}}

\def\spa#1.#2{\left\langle#1\,#2\right\rangle}
\def\spb#1.#2{\left[#1\,#2\right]}
\def\lor#1.#2{\left(#1\,#2\right)}
\def\sand#1.#2.#3{%
\left\langle\smash{#1}{\vphantom1}^{-}\right|{#2}%
\left|\smash{#3}{\vphantom1}^{-}\right\rangle}

\def\cE{\mathcal{E}}

\newcommand{\nn}{\nonumber}

\begin{document}

\title{Quantum Interference in Jet Substructure from Spinning Gluons}
\author{Hao~Chen}
\email{chenhao201224@zju.edu.cn}
\affiliation{Zhejiang Institute of Modern Physics, Department of
  Physics, Zhejiang University, Hangzhou, 310027, China\vspace{0.5ex}}
\author{Ian~Moult}
\email{imoult@slac.stanford.edu}
\affiliation{SLAC National Accelerator Laboratory, Stanford University, CA, 94309, USA\vspace{0.5ex}}
\author{Hua~Xing~Zhu}
\email{zhuhx@zju.edu.cn}
\affiliation{Zhejiang Institute of Modern Physics, Department of
  Physics, Zhejiang University, Hangzhou, 310027, China\vspace{0.5ex}}

\begin{abstract}
Collimated sprays of hadrons, called jets, are an emergent phenomenon of Quantum Chromodynamics~(QCD) at collider experiments, whose detailed internal structure encodes valuable information about the interactions of high energy quarks and gluons, and their confinement into color-neutral hadrons.  The flow of energy within jets is characterized by correlation functions of energy flow operators, with the three-point correlator, being the first correlator with non-trivial shape dependence, playing a special role in unravelling the dynamics of QCD.
In this Letter we initiate a study of the three-point energy correlator to all orders in the strong coupling constant, in the limit where two of the detectors are squeezed together. 
We show that by rotating the two squeezed detectors with respect to the third by an angle $\phi$, a $\cos (2\phi)$ dependence arising from the quantum interference between intermediate virtual gluons with $+/-$ helicity is imprinted on the detector. This can be regarded as a double slit experiment performed with jet substructure, and it provides a direct probe of the ultimately quantum nature of the substructure of jets, and of transverse spin physics in QCD.
To facilitate our all-orders analysis, we adopt the Operator Product Expansion~(OPE) for light-ray operators in conformal field theory and develop it in QCD. Our application of the light-ray OPE in real world QCD establishes it as a powerful theoretical tool with broad applications for the study of jet substructure. 
\end{abstract}

\maketitle


\emph{Introduction.}---Jet substructure, which was originally developed to exploit the flow of energy within jets of particles at the Large Hadron Collider (LHC) to enhance new physics searches  \cite{Butterworth:2008iy}, has since emerged as a primary technique for studying QCD.  By focusing on the internal structure of jets, one obtains a clean probe of the dynamics of QCD on the light cone, which is measured with exquisite angular precision by the LHC detectors.  This has led to a wealth of theoretical and experimental progress with the goal of fully exploiting this remarkable data set \cite{Larkoski:2017jix,Asquith:2018igt,Marzani:2019hun}.

The flow of energy within jets is characterized by correlation functions, $\langle \mathcal{E}(\hat n_1) \mathcal{E}(\hat n_2) \cdots \mathcal{E}(\hat n_k) \rangle$, of light-ray operators
\cite{Sveshnikov:1995vi,Tkachov:1995kk,Korchemsky:1999kt,Bauer:2008dt,Hofman:2008ar,Belitsky:2013xxa,Belitsky:2013bja,Kravchuk:2018htv}
\begin{align}\label{eq:ANEC_op}
\mathcal{E}(\hat n) = \lim_{r\to \infty}  \int\limits_0^\infty dt~ r^2 n^i T_{0i}(t,r \hat n)\,,
\end{align}
in the limit where the $\hat n_i$ are nearly collinear, such that they lie within a single jet. Here $T_{\mu \nu}$ is the energy-momentum tensor, and $\hat n_i$ are unit vectors representing points (calorimeter cells) on the celestial sphere.  Any observable measurable with detectors at infinity can be extracted from the complete set of such correlators \cite{Sveshnikov:1995vi,Tkachov:1995kk}.

The simplest observable is the two-point correlator \cite{Basham:1978bw,Basham:1978zq}, $\langle \mathcal{E}(\hat n_1) \mathcal{E}(\hat n_2) \rangle$, which has no shape dependence. In the collinear limit, the two-point correlator exhibits a scaling behavior \cite{Konishi:1979cb,Richards:1982te,Hofman:2008ar,Kologlu:2019mfz,Korchemsky:2019nzm,Dixon:2019uzg,Chen:2020vvp}, $\langle \mathcal{E}(\hat n_1) \mathcal{E}(\hat n_2)\rangle  \sim 1/\theta_{12}^{2 - 2\gamma(\alpha_s)}$. This scaling probes the value of the strong coupling constant, $\alpha_s$, and has been computed to next-to-next-to-leading logarithm in QCD \cite{Dixon:2019uzg}. 

Much more non-trivial information about the dynamics of QCD is encoded in higher-point correlation functions. In this Letter we show that the squeezed limit of the three-point correlator~\cite{Chen:2019bpb}, $\langle \mathcal{E}(\hat n_1) \mathcal{E}(\hat n_2) \mathcal{E}(\hat n_3) \rangle$, where two detectors are brought together, $\hat n_2 \to \hat n_3$, is a direct probe of quantum interference between helicity $\lambda=\pm$ gluons in the jet, i.e. the transverse spin structure of QCD, and that this is imprinted on the detector as a $\cos(2\phi)$ interference pattern. While the study of transverse spin  in QCD has a long history (see e.g. \cite{Efremov:1981sh,Efremov:1984ip,Qiu:1991pp,Collins:1992kk,Collins:1993kq,Collins:2002kn,Airapetian:2004tw,Boer:1997mf,Qiu:1998ia,Seidl:2008xc,Barone:2010zz,Collins:2011zzd,Kang:2015msa,Kang:2020xyq}), the squeezed limit of the three-point correlator provides a qualitatively new observable for studying transverse spin. Unlike standard probes of transverse spin that rely on non-perturbative fragmentation processes, energy correlators are infrared and collinear safe, and for the high energy jets at the LHC, exhibit a regime where the transverse spin structure can be computed in perturbative QCD. In the extreme collinear limit they are sensitive to non-perturbative fragmentation, allowing for the study of the transition between perturbative and non-perturbative physics. Furthermore, they do not involve polarized beams or hadrons, or reference directions, and so are ideally suited for the LHC. 

\begin{figure*}[ht!]
  \centering
  \qquad   \subfloat[]{\includegraphics[width=0.29\textwidth]{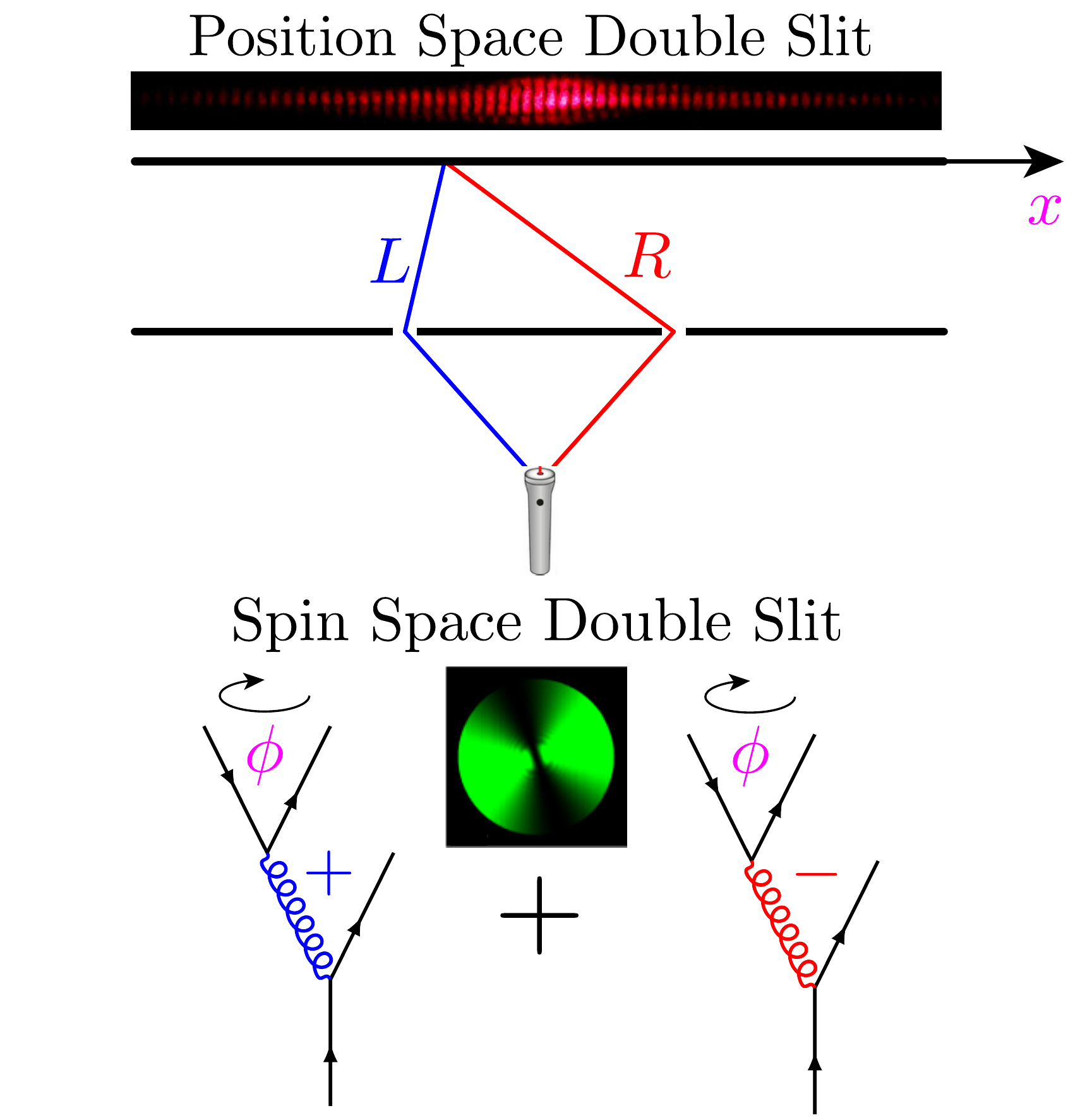}}\qquad\qquad \qquad
    \subfloat[]{\includegraphics[width=0.44\textwidth]{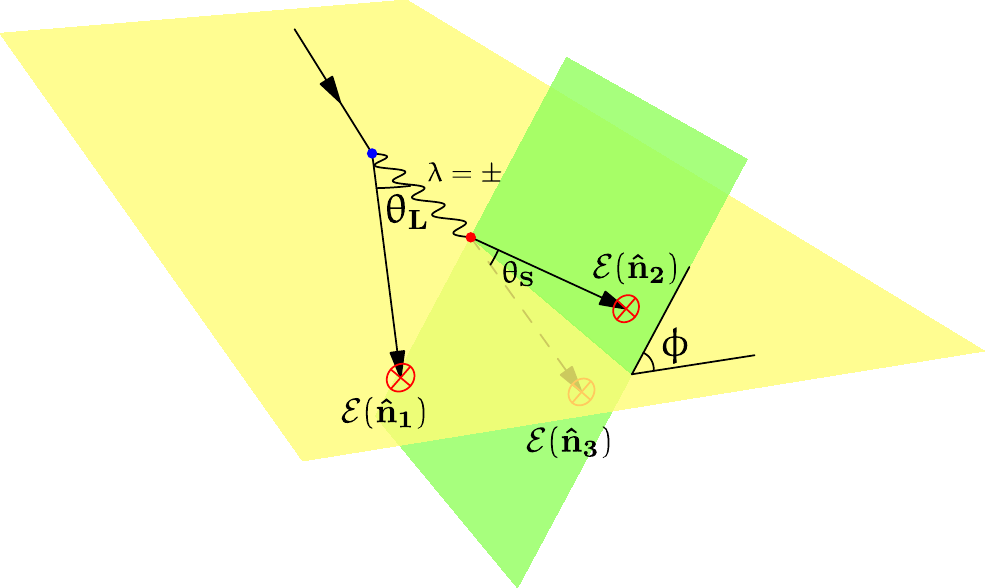}}
  \caption{Double slit experiments in position and spin space (a), and physical implementation of the spin space double slit experiment using the squeezed limit, $\theta_S \ll \theta_L$, of the three-point correlator (b). Quantum interference between gluon spin states, $\lambda=\pm$, leads to a $\cos(2\phi)$ pattern as the squeezed correlators are rotated.}
  \label{fig:slit}
\end{figure*}

The description of the complicated all orders perturbative structure of multi-point correlators calls for the application of new theoretical tools. Much in analogy with how the local operator product expansion \cite{Wilson:1969zs}, $\cO(x_1) \cO(x_2)= \sum (x_2-x_1)^{\gamma_i} \cO_i(x_1)$, describes the behavior of local operators as they are brought together, in the limit of interest in jet substructure where the light-ray operators $\mathcal{E}(\hat n)$ become collinear, there exists a recently developed light-ray OPE \cite{Hofman:2008ar,Kravchuk:2018htv,Kologlu:2019bco,Kologlu:2019mfz,1822249} $\mathcal{E}(\hat n_1) \mathcal{E}(\hat n_2)=\theta^{\gamma_i} \sum\mathbb{O}_i(\hat n_1)$, where the $\mathbb{O}_i(\hat n)$ are non-local light-ray operators \cite{Balitsky:1987bk,Balitsky:1988fi,Balitsky:1990ck,Braun:2003rp,Hofman:2008ar,Kravchuk:2018htv}.
However, the use of the light-ray OPE has so far been restricted to conformal field theories (CFTs) \cite{Hofman:2008ar,1822249,Kologlu:2019mfz,Kologlu:2019bco}.

In this Letter we show that the light-ray OPE can be applied in QCD, and  provides powerful operator based techniques for jet substructure. We show that the iterated OPE of $\mathcal{E}(\hat n)$ operators closes at leading twist onto operators $\mathbb{O}^{[J]}_i(\hat n)$ with arbitrary collinear spin-$J$, but restricted transverse spin-$j=0,2$, and  we explicitly compute the $\mathcal{E}(\hat n_1) \mathcal{E}(\hat n_2)$ and $\mathbb{O}^{[J]}_i (\hat n_1)  \mathcal{E}(\hat n_2)$ OPEs.
The all orders structure of spin interference effects in the three-point correlator then arises naturally from the transverse spin structure of the light-ray OPE.


\emph{Interference in the Squeezed Limit.}---The physics of the squeezed limit of the three-point correlator in a weakly coupled gauge theory can be described as a double slit experiment in spin space, see Fig.~\ref{fig:slit}. The interference pattern in the usual double slit experiment is due to the interference in $|A_L(x) + A_R(x)|^2$, where $A_{L(R)}(x)$ is the amplitude for going through the left~(right) slit from the light source to position $x$ on the detector. Similarly, in the squeezed limit of the three-point correlator, the interference terms in  $|A_+(\phi) + A_-(\phi)|^2$ are the source of an interference pattern, where $A_{+(-)}$ is the splitting amplitude with a nearly on-shell virtual gluon with positive~(negative) helicity. Therefore the slits in the standard double slit experiment are replaced by the intermediate $+/-$ helicity gluons, and varying the distance $x$ is replaced by varying the angle $\phi$ of the squeezed energy correlators. We emphasize that while this effect arises from quantum interference, we have been unable to prove a Bell-type inequality using only energy measurements. It would be interesting to understand if Bell-type inequalities can be proven in the collider context, even in principle. Similar questions have also been considered in the context of inflationary measurements \cite{Maldacena:2015bha}.

We parametrize the squeezed limit symmetrically, using $(\theta_S,\theta_L, \phi)$ as shown in Fig.~\ref{fig:slit}, to eliminate linear power corrections in $\theta_S/\theta_L$. The squeezed limit is characterized by $\theta_S\ll \theta_L$, with $\phi$ arbitrary, and the expansion in this limit takes the form
\begin{align}
  \label{eq:expansion}
      \frac{d^3 \Sigma}{d\theta_L^2 d\theta_S^2 d\phi} \simeq  \frac{1}{\pi}  \left(\frac{\alpha_s}{4 \pi} \right)^2   \frac{ \text{Sq}_i^{(0)}(\phi)}{\theta_L^2 \theta_S^2} +\cdots \,,
\end{align}
where the dots denote terms less singular in the squeezed limit. Expanding the full result for the three-point correlator in \cite{Chen:2019bpb}, we find for quark and gluon jets, 
\begin{align}\label{eq:fixed_order}
\text{Sq}^{(0)}_q (\phi)&=C_F n_f T_F \left(  \frac{39-{\color{blue}20\cos(2\phi)}}{225} \right) \\
                 &+C_F C_A \left(  \frac{273+{\color{red}10\cos(2\phi)}}{225} \right)+C_F^2 \frac{16}{5}
                   \nn\\
  &\hspace{-1cm} = 10.54+0.1156 n_f+(0.1778-0.0593 n_f) \cos(2\phi),\nn \\
\text{Sq}^{(0)}_g (\phi) &=C_A n_f T_F  \left(  \frac{126-{\color{blue}20\cos(2\phi)}}{225} \right)\nn \\
&+C_A^2 \left(  \frac{882+{\color{red}10\cos(2\phi)}}{225} \right)+C_Fn_f T_F \frac{3}{5}\nn\\
&\hspace{-1cm}  =(35.28+1.24 n_f)+ (0.4-0.133 n_f)\cos(2\phi)\,. \nn
\end{align}
Here we see $\cos(2\phi)$ interference terms at leading twist, which at this order are identical for quark and gluon jets, since they arise only from an intermediate gluon, and  have opposite signs for $g\to q\bar q$ (in blue) and $g\to gg$ (in red). Positivity of the cross section guarantees that the $\cos(2\phi)$ terms are smaller than the constant terms, analogous to the conformal collider bounds \cite{Hofman:2008ar}. Due to the singular structure of the squeezed limit, the all orders resummation of these spin interference effects is required to describe the three-point correlator, as well as for limits of higher-point correlators.

Despite their importance for observables relevant to jet substructure, spin interference effects are not included in the standard parton shower simulations used to this point by experiments at the LHC, which, as illustrated through detailed studies in \cite{Dasgupta:2020fwr}, can lead to large errors for multi-emission jet substructure observables. Recently, significant progress was made with the implementation of algorithms for including spin interference effects in parton showers \cite{Collins:1987cp,Knowles:1988hu,Knowles:1988vs,Knowles:1987cu,Richardson:2001df,Richardson:2018pvo} in a new version of Herwig \cite{Richardson:2018pvo,Bellm:2019zci}. The goal of this Letter will be to understand the all orders structure of these spin interference effects analytically, and show how they emerge in a simple manner from the transverse spin structure of the light-ray OPE.

Transverse spin can also be studied using the two-point correlator with a polarized source \cite{1822249}. However, for a source coupling to quarks, polarization effects appear first at $\cO(\alpha_s^2)$. While the unpolarized cross section is known at this order in QCD \cite{Dixon:2018qgp,Luo:2019nig} and beyond in $\cN=4$ super-Yang-Mills (SYM) \cite{Belitsky:2013ofa,Henn:2019gkr}, the polarized cross-section is not (or is related by superconformal symmetry to the scalar case \cite{Belitsky:2014zha,Korchemsky:2015ssa,1822249}). Spin interference effects should appear at $\cO(\alpha_s)$ for the two-point correlator in QCD with the stress tensor as a source, which would be interesting to calculate.


\emph{The Light-Ray OPE in QCD.}---To our knowledge, the analytic resummation of spin interference effects in the collinear limit has not been achieved using standard perturbative QCD techniques. This is in contrast to the more well studied Sudakov (or back-to-back) region (see e.g. \cite{Chien:2020hzh,Hatta:2020bgy}).
To achieve this resummation, we will apply the light-ray OPE formalism \cite{Hofman:2008ar,Kravchuk:2018htv,Kologlu:2019bco,Kologlu:2019mfz,1822249}.  In addition to this particular phenomenological application, our calculation provides a highly non-trivial application of the light-ray OPE in a non-conformal field theory, and illustrates its potential for jet substructure at the LHC.

\begin{figure}
\includegraphics[width=1.0\linewidth]{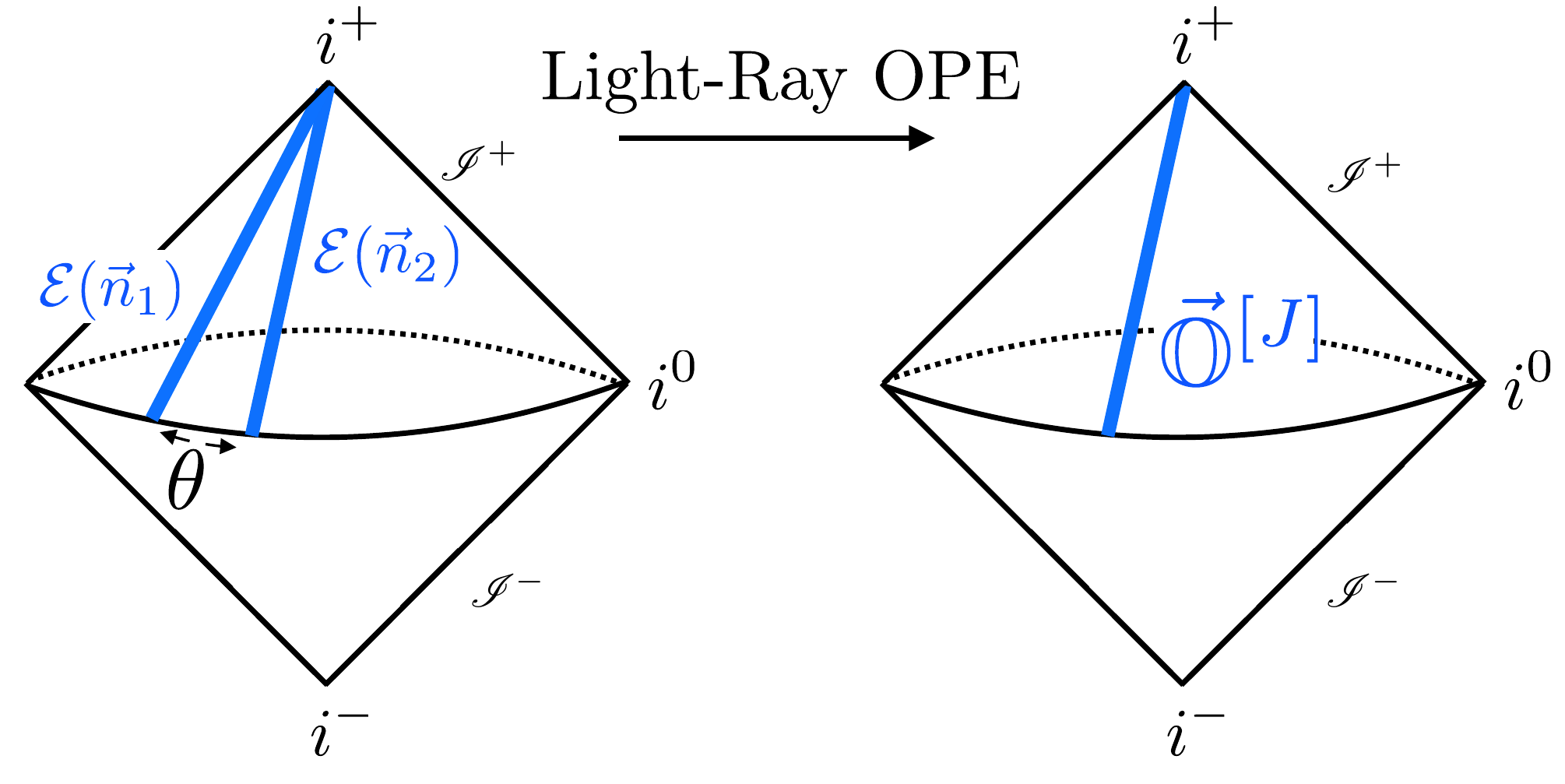}
\caption{At leading twist, the light-ray OPE in QCD for $\mathcal{E}$ operators leads to the light-ray operators $\vec{\mathbb{O}}^{[J]}$ characterized by a spin $J$, and transverse spin $j=0,2$.}
\label{fig:OPE}
\end{figure}

The OPE of light-ray operators was originally presented \cite{Hofman:2008ar} in the context of gauge theories using the string operators of \cite{Balitsky:1987bk}. It has recently been developed into a rigorous OPE in CFTs \cite{Kravchuk:2018htv,Kologlu:2019bco,Kologlu:2019mfz,1822249}. Although the loss of symmetries relaxes the rigid structure of the OPE in a CFT, the intuition for its existence remains. Furthermore, it is well known that QCD often inherits structure from the conformal limit \cite{Braun:2003rp}.

For multiple light-ray operators with small transverse separations, one expects an OPE onto a sum of light-ray operators each formed from potentially multiple operators smeared on the light cone, as shown in the Penrose diagram in \Fig{fig:OPE}.  Characterizing the operators involved by their twist, collinear spin-$J$ and transverse spin-$j$, in a weakly coupled gauge theory, such as QCD in the perturbative regime, we have the simplification that the leading twist contributions involve two fields, and therefore can only have transverse spin-$0$ or $2$. This greatly simplifies the OPE, allowing it to be iterated to any number of light-ray operators. The structure beyond leading twist is known for the $\cE(\hat n_1) \cE(\hat n_2)$ OPE in CFTs \cite{1822249}.

The leading twist operators in QCD are (see e.g. \cite{Christ:1972ms})
\begin{align}
\cO_q^{[J]}&=\frac{1}{2^J}\bar \psi \gamma^+ ( iD^+)^{J-1}\psi\,, \nn \\
\cO_g^{[J]}&=-\frac{1}{2^J} F_a^{\mu +}(i D^+)^{J-2}F_a^{\mu +}\,, \nn \\
\cO_{\tilde g,\lambda}^{[J]}&=-\frac{1}{2^J} F_a^{\mu +}(i D^+)^{J-2}F_a^{\nu +}\epsilon_{\lambda, \mu}\epsilon_{\lambda, \nu}\,,
\end{align}
where the plus component of a vector $A^\mu$ is $\bar{n} \cdot A$, and $\bar{n} = (1, - \hat{n})$.
We have projected the transverse spin-$2$ operators onto helicities $\lambda=\pm$ using polarization vectors, $\epsilon^\mu$. These transverse spin-$2$ operators are purely gluonic, and will reproduce the interference terms in Eq.~\ref{eq:fixed_order}.  Smearing on the light cone (the light transform \cite{Kravchuk:2018htv}) gives rise to light-ray operators, which we write as a vector
\begin{align}
\vec{\mathbb{O}}^{[J]}=\left( \mathbb{O}_q^{[J]}, \mathbb{O}_g^{[J]}, \mathbb{O}_{\tilde g,+}^{[J]}, \mathbb{O}_{\tilde g,-}^{[J]}  \right)^T=\lim_{r\to \infty} r^2 \int\limits_0^{\infty} dt\, \vec{\cO}^{[J]}(t,r\vec{n})\,. \nn
\end{align}

To derive the leading twist $\cE(\hat n_1) \cE(\hat n_2)$ OPE in the leading logarithmic approximation, we perform an explicit matching calculation using Wightman functions. Since the OPE is state independent, we compute the OPE coefficients by performing a matching calculation using states produced by gauge invariant collinear quark or gluon fields. At lowest order in perturbation theory the diagrams entering the matching calculation describe the splitting of the quark or gluon field into the two energy detectors. We find
\begin{align}\label{eq:EE_OPE}
\cE(\hat n_1) \cE(\hat n_2)& = -\frac{1}{2\pi} \frac{2}{\theta^2}  \vec {\cal J} \left[  \widehat C_\phi(2)- \widehat C_\phi(3)  \right] \vec {\mathbb{O}}^{[3]} (\hat n_1)\nn \\
&+ \text{ higher twist}\,,
\end{align}
where $\theta$ is the angle between the OPEd pair, and $\vec {\cal J}=(1,1,0,0)$ is a projector onto unpolarized quark and gluon twist operators, and the matrix of OPE coefficients is 
\begin{widetext}
\begin{align}\label{eq:structure_constants}
\widehat C_\phi(J)&=
\begin{pmatrix}
\gamma_{qq}(J)&&2n_f \gamma_{qg}(J)&&2n_f \gamma_{q\tilde g}(J) e^{-2i\phi}/2 &&2n_f \gamma_{q\tilde g}(J) e^{2i\phi}/2 \\
\gamma_{gq}(J)&& \gamma_{gg}(J)&&\gamma_{g\tilde g}(J) e^{-2i\phi}/2 &&\gamma_{g\tilde g}(J) e^{2i\phi}/2\\
\gamma_{\tilde gq}(J) e^{2i\phi} &&\gamma_{\tilde gg}(J) e^{2i\phi} && \gamma_{\tilde g \tilde g}(J)&&\gamma_{\tilde g \tilde g,\pm}(J) e^{4i\phi}\\
\gamma_{\tilde gq}(J) e^{-2i\phi} &&\gamma_{\tilde gg}(J) e^{-2i\phi} &&\gamma_{\tilde g \tilde g,\pm}(J)e^{-4i\phi}&& \gamma_{\tilde g \tilde g}(J)\\
\end{pmatrix}\,.
\end{align}
\end{widetext}
The quantum number $J=3$ appearing in  Eq.~\ref{eq:EE_OPE} arises from the gauge theory perspective due to the fact that the two detectors lead to an $E^2$ weighting of the splitting function, corresponding to its third moment \cite{Dixon:2019uzg}, while from a CFT perspective, it is fixed by the quantum numbers of the light-ray operators \cite{Hofman:2008ar}.

While the general structure of this OPE was anticipated in \cite{Hofman:2008ar,1822249}, the explicit form of the OPE in QCD is a new result. An interesting feature of Eq.~\ref{eq:structure_constants} is that the OPE coefficients of light-ray operators are themselves anomalous dimensions. Indeed, $\gamma_{ij}(J)$, $i,j=q,g$ are the standard twist-2 spin-$J$ anomalous dimensions, related to the unpolarized splitting functions, while $\gamma_{i,\tilde g}(J)$, $i=q,g$ are moments of the spin correlated splitting functions of \cite{deFlorian:2000pr}. This builds a direct bridge between the language of splitting functions in perturbative gauge theories and the light-ray OPE, which deserves further exploration. On the other hand, $\gamma_{\tilde g, i}(J)$ have,  to our knowledge, not appeared previously in the QCD literature. They do not appear in the $\cE(\hat n_1) \cE(\hat n_2)$ OPE due to the projector $\vec {\cal J}$, but will appear in the more general $\mathbb{O}^{[J]}(\hat n_1)  \mathcal{E}(\hat n_2)$ OPE considered below.  Denoting the perturbative expansion as $\gamma_{ij}=(\alpha_s/(4\pi))\gamma_{ij}^{(0)}+\cO(\alpha_s^2)$, we have
\begin{align}\label{eq:structure_explicit}
\gamma^{(0)}_{\tilde g \tilde g} (J)&=4C_A (\psi ^{(0)}(J+1)+\gamma_E) -\beta_0\,,  \nn\\
  \gamma^{(0)}_{\tilde gq}(J)&=C_F\frac{2}{(J-1)J}\,,\qquad  \gamma^{(0)}_{\tilde gg}(J)=C_A\frac{2}{(J-1)J}\,,\nn \\
\gamma^{(0)}_{q\tilde g}(J)&= - T_F\frac{8}{(J+1)(J+2)}\,, \qquad \gamma^{(0)}_{\tilde g \tilde g,\pm} (J)=0\,, \nn \\
  \gamma^{(0)}_{g\tilde g}(J)&= C_A \left(\frac{8}{(J+1)(J+2)}+3\right) -\beta_0\,,
\end{align}
where $\psi^{(0)}$ is the Digamma function, and $\beta_0 = 11 C_A/3 - 4 T_F n_f/3$ is the first order QCD $\beta$-function.
The other anomalous dimensions are standard (see e.g. \cite{Chen:2020vvp}).

Plugging the explicit values of the OPE coefficients in Eq.~\ref{eq:structure_constants} for $J=3$ into the $\cE(\hat n_1) \cE(\hat n_2)$ OPE of  Eq.~\ref{eq:EE_OPE} reproduces the result for the two-point correlator in QCD for both quark and gluon sources \cite{Dixon:2019uzg}, providing a non-trivial check of Eq.~\ref{eq:EE_OPE}.

In a CFT, the collinear spin $J=3$ appearing in the $\cE(\hat n_1) \cE(\hat n_2)$ OPE is fixed by symmetry \cite{Hofman:2008ar,Kologlu:2019mfz}. From the structure of higher logarithmic terms in \cite{Dixon:2019uzg}, we expect  that in QCD light-ray operators with $J=3+\cO(\alpha_s)$ appear at higher orders due to the breaking of conformal symmetry and the non-trivial manifestation of Basso-Korchemsky reciprocity \cite{Basso:2006nk,Chen:2020uvt} in the EEC \cite{Dixon:2019uzg}.


\emph{Squeezed Limits from Light-Ray OPEs.}---To compute the squeezed limit of the three-point correlator, $\langle \mathcal{E}(\hat n_1) \mathcal{E}(\hat n_2) \mathcal{E}(\hat n_3) \rangle$, we perform the successive OPE $\mathcal{E}(\hat n_2) \mathcal{E}(\hat n_3) \to \sum \mathbb{O}^{[3]}(\hat n_2)$ followed by $\mathcal{E}(\hat n_1)  \vec{\mathbb{O}}^{[J]}(\hat n_2)  \to \vec{\mathbb{O}}^{[J+1]} (\hat n_1)$. The second OPE is computed analogously to the $\mathcal{E}(\hat n_1) \mathcal{E}(\hat n_2)$ OPE discussed above. Due to the fact that the maximal transverse spin at leading twist in QCD is $2$, we find that the iterated OPE of light-ray operators closes onto the operators $\vec {\mathbb{O}}^{[J]}(\hat n_1)$. To be able to apply our results to arbitrary iterated squeezed limits, we compute the OPE coefficients in the $  \mathcal{E}(\hat n_1)  \vec{\mathbb{O}}^{[J]}  (\hat n_2)$ OPE as analytic functions of the collinear spin-$J$. 

The $\vec {\mathbb{O}}^{[J]} (\hat n_1) \cE(\hat n_2) $ OPE is
\begin{align}\label{eq:OE_OPE}
\vec {\mathbb{O}}^{[J]} (\hat n_1) \cE(\hat n_2)     &=-\frac{1}{2\pi}\frac{2}{\theta^2}\left[  \widehat C_\phi(J)- \widehat C_\phi(J+1)  \right] \vec {\mathbb{O}}^{[J+1]} (\hat n_1)\nn \\
&+\text{higher twist}\,.
\end{align}
Unlike the $\cE(\hat n_1) \cE(\hat n_2)$ OPE this involves the full matrix of OPE coefficients including the entries $\gamma_{\tilde g, i}(J)$.

In the squeezed limit, the hierarchy between the splitting angles, $\theta_S\ll \theta_L$, gives rise to large logarithmic corrections in OPE coefficients that must be resummed to all orders to obtain a physical result. This resummation is performed by solving the renormalization group equations for the light-ray operators, 
\begin{align}\label{eq:resum}
\frac{d}{d\ln \mu^2} \vec {\mathbb{O}}^{[J]} = - \hat \gamma(J) \cdot  \vec {\mathbb{O}}^{[J]}\,, 
\end{align}
where
\begin{align}\label{eq:anom_dim}
\hat \gamma(J)=
\begin{pmatrix}
\gamma_{qq}(J)&&2n_f \gamma_{qg}(J)&&0\\
\gamma_{gq}(J)&& \gamma_{gg}(J)&&0\\
0&&0&& \gamma_{\tilde g \tilde g}(J)\mathbf{1}
\end{pmatrix}\,,
\end{align}
with $\mathbf{1}$ a $2\times 2$ identity matrix.
Combining Eqs.~\ref{eq:EE_OPE}, \ref{eq:OE_OPE} and \ref{eq:resum}, we derive a resummed result describing the leading twist singular behavior in the squeezed limit of the three-point correlator at leading-logarithmic order
\begin{widetext}
\begin{align}\label{eq:final_formula}
\mathcal{E}(\hat n_1) \mathcal{E}(\hat n_2) \mathcal{E}(\hat n_3)=\frac{1}{(2\pi)^2} \frac{2}{\theta_S^2}\frac{2}{\theta_L^2} \vec {\cal J} \left[ \widehat C_{\phi_S}(2)- \widehat C_{\phi_S}(3)  \right] \left[ \frac{\alpha_s(\theta_L Q)}{\alpha_s(\theta_S Q)}\right]^{\frac{\widehat \gamma^{(0)}(3)}{\beta_0}} \left[ \widehat C_{\phi_L}(3)-\widehat C_{\phi_L}(4)  \right]  \left[ \frac{\alpha_s(Q)}{\alpha_s(\theta_L Q)}\right]^{\frac{\widehat \gamma^{(0)}(4)}{\beta_0}} 
\vec {\mathbb{O}}^{[4]} (\hat n_1) 
+\cdots
\end{align}
\end{widetext}
where the azimuthal angle $\phi$ in Eq.~\ref{eq:expansion} is identified as $\phi_S - \phi_L$, and the overall rotation of the jet can be integrated out since we consider unpolarized sources. 
The dots denote higher twist and subleading logarithmic contributions.
Plugging in the explicit values for the anomalous dimensions, Eq.~\ref{eq:structure_explicit}, and expanding to leading order in $\alpha_s$, we find that Eq.~\ref{eq:final_formula} exactly reproduces the fixed order results in Eq.~\ref{eq:fixed_order}, providing a highly non-trivial test of our OPE formulas.

We can also consider the squeezed limit of the three-point correlator in $\cN=4$ SYM  \cite{Chen:2019bpb}. Here, the evolution matrix in Eq.~\ref{eq:anom_dim}, reduces to a scalar evolution with a universal anomalous dimension \cite{Kotikov:2004er}, and $\gamma_{\tilde g \tilde g}=\gamma_{gg}$ \cite{Beisert:2003jj,Hofman:2008ar}, so Eq.~\ref{eq:final_formula} agrees with the prediction for the scaling of the transverse spin-$0$ and spin-$2$ contributions from \cite{Hofman:2008ar}.  However, we find that the leading twist spin correlations vanish after summing over the multiplet, since $\sum [C_\phi(3)]_{i,\tilde g}=0$, $i=\phi, q,g$. This agrees with the perturbative prediction of \cite{Chen:2019bpb}. This is one manifestation of the ``classicality'' of $\cN=4$ SYM.

\begin{figure}[t!]
 \hspace{-0.2cm}   \subfloat[]{\includegraphics[width=0.30\textwidth]{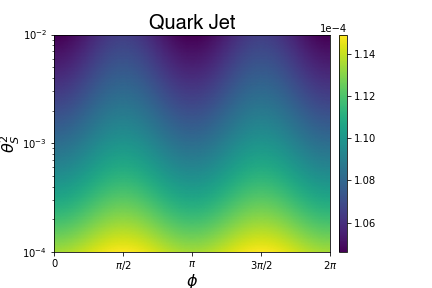}}
    \subfloat[]{\includegraphics[width=0.30\textwidth]{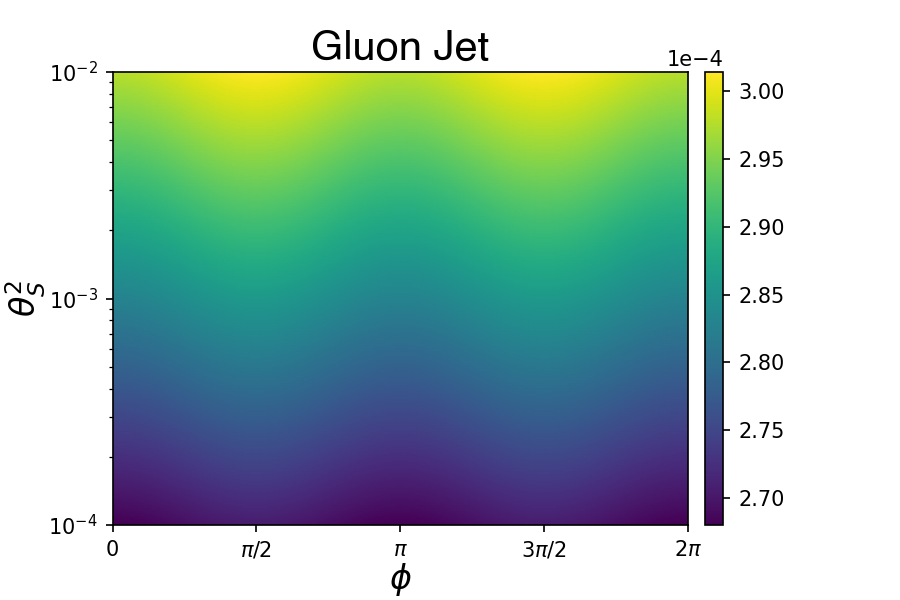}}
\\    \subfloat[]{\includegraphics[width=0.30\textwidth]{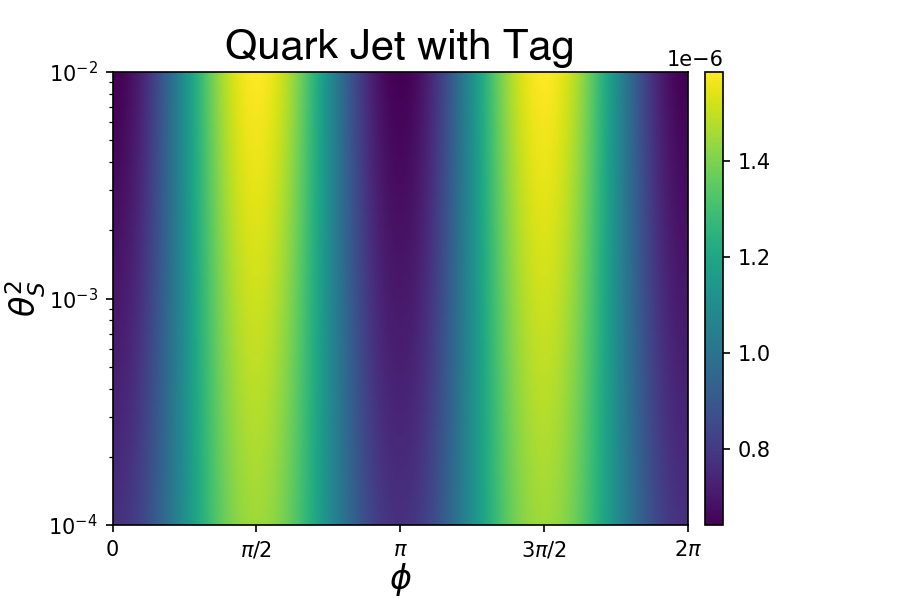}}    \subfloat[]{\includegraphics[width=0.30\textwidth]{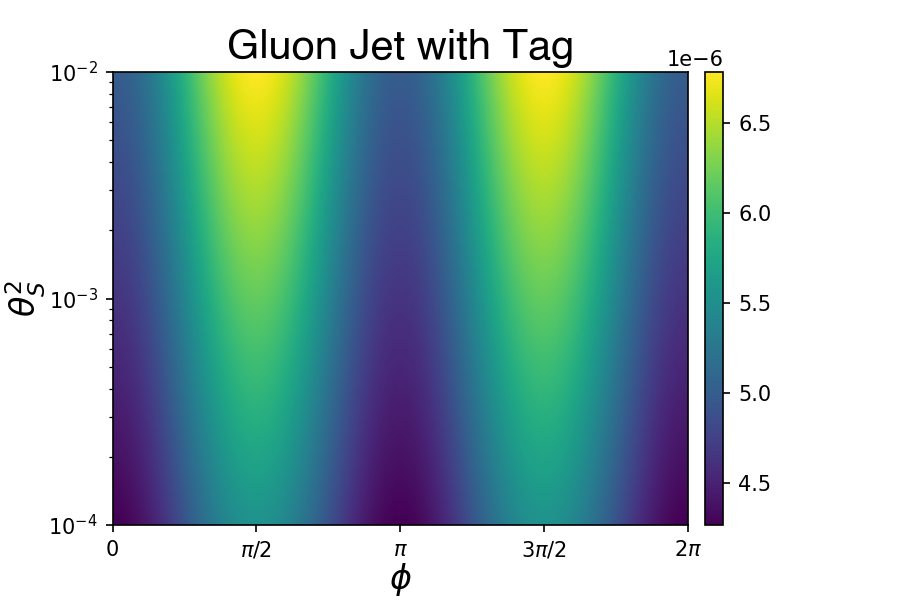}}
\caption{The squeezed limit of the three-point correlator for (a) quark and (b) gluon jets at the LHC. Interference effects are seen as the $\cos(2\phi)$ ripple modulated by resummation in $\theta_S$. In (c), (d), we assume an idealized b-tagger, as explained in the text, to enhance the effect.
}
\label{fig:LHC}
\end{figure}

\emph{Numerical Results at the LHC.}---Using Eq.~\ref{eq:final_formula}, we make numerical predictions for unpolarized quark and gluon jets at the LHC. In \Fig{fig:LHC} (a) and (b), we show the squeezed limit of the three-point correlator, weighted by $\theta_L^2 \theta_S^2$, as a function of $(\phi, \theta_S)$ for fixed $\theta_L$. Here we have used $Q=1$ TeV,  $\theta_L^2=0.1$ and $\alpha_s(1$TeV$)=0.087$, as well as the one-loop $\beta$-function. The ripples in the distribution are clearly visible, and illustrate the direct imprint of quantum interference effects in the detector. They are modulated by the resummation in $\theta_S$, which has a qualitatively different structure for quark and gluon jets, as discussed in \cite{Dixon:2019uzg}.

For the case of QCD with $n_f=5$ light flavors, the interference effects are at the few percent level due to a cancellation of $g\to q\bar q$ and $g\to gg$ splittings. However, we believe that if measured using tracks, they are on the boundary of what can be achieved (see e.g. \cite{Aad:2019vyi}). Furthermore, there are a number of ways to enhance the interference signals, including using charge information to identify the $g\to qq$ splitting, or $b$-tagging to perform the measurement. In \Fig{fig:LHC} (c) and (d), we show predictions using an idealized b-tagging on the squeezed pair, which isolates the $g\to bb$ contribution, and enhances the modulation to an $\cO(1)$ effect. We leave a detailed phenomenological study of the optimal strategy to future work.


\emph{Conclusions.}---In this Letter we have studied the three-point energy correlator, $\langle \mathcal{E}(\hat n_1) \mathcal{E}(\hat n_2) \mathcal{E}(\hat n_3) \rangle$, at the LHC. Our study is novel both phenomenologically, where we have proposed to use squeezed limits of energy correlators to probe quantum interference and transverse spin effects in jet substructure, as well as theoretically, where we have developed the light-ray OPE in  QCD, and showed that it provides a transparent way of understanding the resummation of spin interference effects.

Our results for the $\cE(\hat n_1) \cE(\hat n_2)$ and  $\cE(\hat n_1) \mathbb{O}^{[J]}(\hat n_2)$ OPEs in QCD allow the description of iterated squeezed limits of $n$-point correlators and opens the door for these observables to be used as precision probes of QCD at the LHC. The measurement of these multi-point event correlators in the deep non-perturbative regime is also fascinating as they are sensitive to polarization effects in the non-perturbative fragmentation process.
 
 
There are numerous avenues for further theoretical development of the light-ray OPE in QCD, including understanding the appearance of light-ray operators with non-integer spin at higher logarithmic orders, the structure of the $1\to 3$ light-ray OPE, and the inclusion of electromagnetically charged correlators \cite{Chicherin:2020azt}. Furthermore, many techniques developed to exploit symmetries in the CFT setting, such as the use of conformal blocks to resum higher twist corrections  \cite{Kologlu:2019mfz,1822249}, or the use of differential operators to relate contributions with different quantum numbers \cite{Karateev:2017jgd}, can be applied in the QCD context to significantly extend standard splitting function based approaches. We therefore believe that the application of the light-ray OPE to problems of phenomenological interest at the LHC can lead to fruitful collaboration between the CFT and jet substructure communities, leading to significant advances in our understanding of QCD.


\emph{Acknowledgements.}---We thank Ming-xing Luo, Tong-Zhi Yang, Xiao-Yuan Zhang for helpful discussions about energy correlators in perturbation theory, David Simmons-Duffin, Petr Kravchuk, and Cyuan-Han Chang for helpful discussions about energy correlators and the light-ray OPE, Jesse Thaler and Patrick Komiske for helpful discussions about aspects of energy correlators at the LHC, and Duff Neill for helpful discussions about the nuclear physics literature. We thank Alexander Karlberg for helpful comments on the manuscript.  H.C. and H.X.Z. are supported by National Natural Science
Foundation of China under contract No. 11975200.
I.M. is supported by the Office of High Energy Physics of the U.S. DOE under Contract No. DE-AC02-76SF00515.

\bibliography{spinning_gluon.bib}{}
\bibliographystyle{apsrev4-1}

\end{document}